\documentclass{article}
\usepackage{ijcai20}

\usepackage{times}
\usepackage{soul}
\usepackage{url}
\usepackage[hidelinks]{hyperref}
\usepackage[utf8]{inputenc}
\usepackage[small]{caption}
\usepackage{graphicx}
\usepackage{amsmath}
\usepackage{amsthm}
\usepackage{booktabs}
\usepackage{algorithm}
\usepackage{algorithmic}
\usepackage{subcaption}
\usepackage{amsmath,amssymb}
\usepackage{graphicx}
\usepackage{array}
\usepackage{enumitem}
\usepackage{multirow}
\newcolumntype{L}[1]{>{\raggedright\let\newline\\\arraybackslash\hspace{0pt}}m{#1}}
\newcolumntype{C}[1]{>{\centering\let\newline\\\arraybackslash\hspace{0pt}}m{#1}}
\newcolumntype{R}[1]{>{\raggedleft\let\newline\\\arraybackslash\hspace{0pt}}m{#1}}

\title{Exploring Effects of Random Walk Based Minibatch Selection Policy on Knowledge Graph Completion}

\author{
Bishal Santra$^1$
\and
Prakhar Sharma$^1$\and
Sumegh Roychowdhury$^1$\And
Pawan Goyal$^1$
\affiliations
$^1$Indian Institute of Technology Kharagpur\\
Kharagpur, India
\emails
\{bsantraigi, prakhar6sharma, sumeghtech, pawang.iitk\}@gmail.com}

\begin{document}

\maketitle

%% ACTUAL PAPER CONTENT
\begin{abstract}
In this paper, we have explored the effects of different minibatch sampling techniques in Knowledge Graph Completion. Knowledge Graph Completion (KGC) or Link Prediction is the task of predicting missing facts in a knowledge graph. KGC models are usually trained using margin, soft-margin or cross-entropy loss function that promotes assigning a higher score or probability for true fact triplets. Minibatch gradient descent is used to optimize these loss functions for training the KGC models. But, as each minibatch consists of only a few randomly sampled triplets from a large knowledge graph, any entity that occurs in a minibatch, occurs only once in most cases. Because of this, these loss functions ignore all other neighbors of any entity, whose embedding is being updated at some minibatch step. In this paper, we propose a new random-walk based minibatch sampling technique for training KGC models that optimizes the loss incurred by a minibatch of closely connected subgraph of triplets instead of randomly selected ones. We have shown results of experiments for different models and datasets with our sampling technique and found that the proposed sampling algorithm has varying effects on these datasets/models. Specifically, we find that our proposed method achieves state-of-the-art performance on the DB100K dataset.
% Our experiments also show that the effect of this sampling technique is consistent across all different models for the KGC task. 
%Also because of the simplicity of the proposed sampling technique, it can be applied in any future models for KGC that might be developed by researchers.

%Our experiments also show that with this loss function many KGC models, starting from linear models like TransE and Distmult to more complex state-of-the-art models like RotatE perform better than their originally published counterparts. 
\end{abstract}

% KGC models can be interpreted in general as a encoder-decoder architecture. The encoder encodes entities and relations in a knowledge graph into fixed dimension vector representations. The decoder assigns probabilities or scores to all fact triplets (two entities and a directed relation edge between them) in the knowledge graph.
% These loss functions for these models are optimized using minibatch gradient descent. 
% Although the loss functions try to maximize the margin between scores assigned to the positive and negative triplets in the graph, this is not fully realized in a minibatch setting, as the minibatch consists of only randomly sampled triplets.
% In this paper, we propose a new family of loss functions for KGC models that optimizes fact score/probability in subgraphs around randomly sampled entities. 

\begin{figure*}[h]
	\centering
	\begin{subfigure}{.19\textwidth}
		\includegraphics[width=0.9\linewidth]{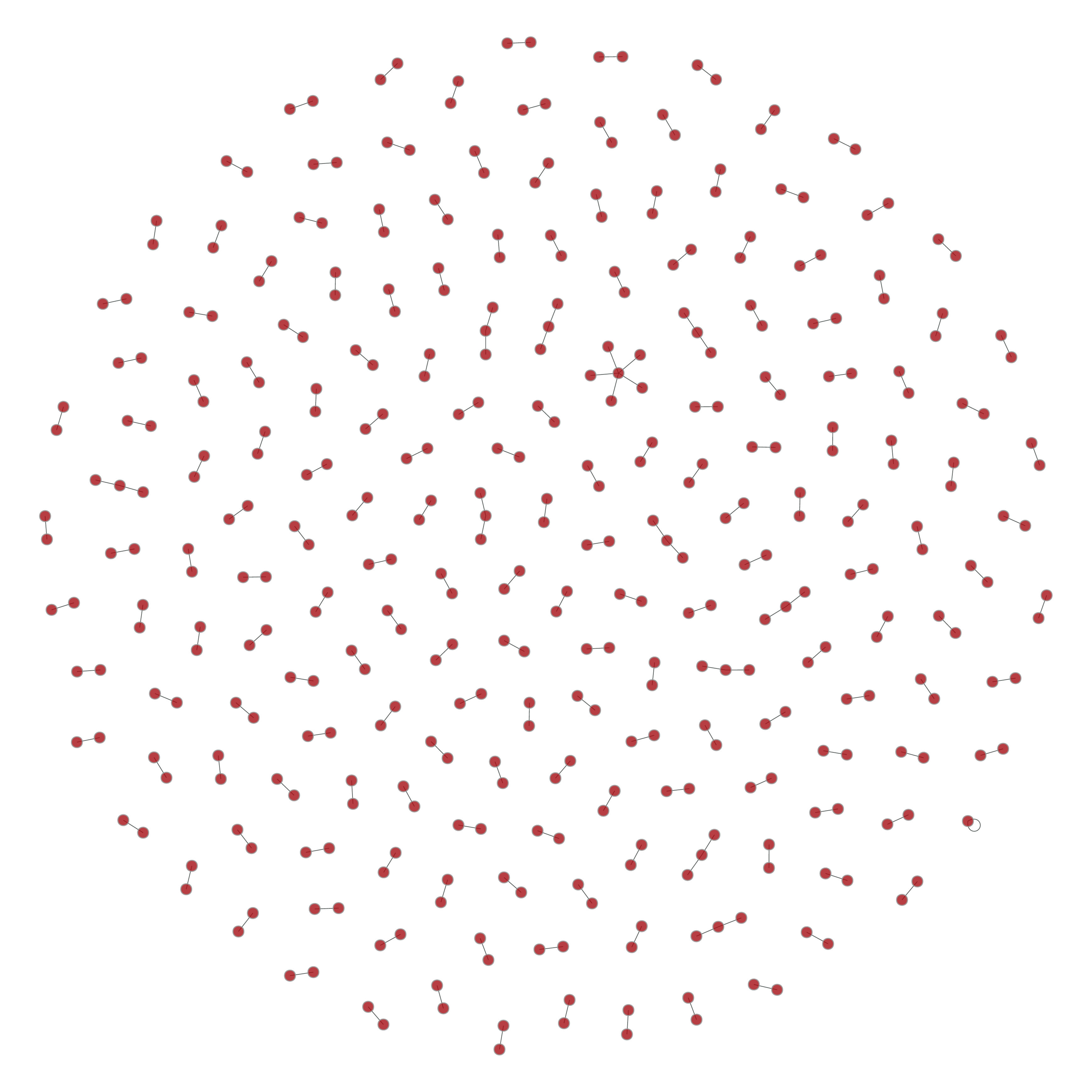}
		\caption{Simply Random}
		\label{fig:graphsrfb15k-237}
	\end{subfigure}
	\begin{subfigure}{.19\textwidth}
		\includegraphics[width=0.9\linewidth]{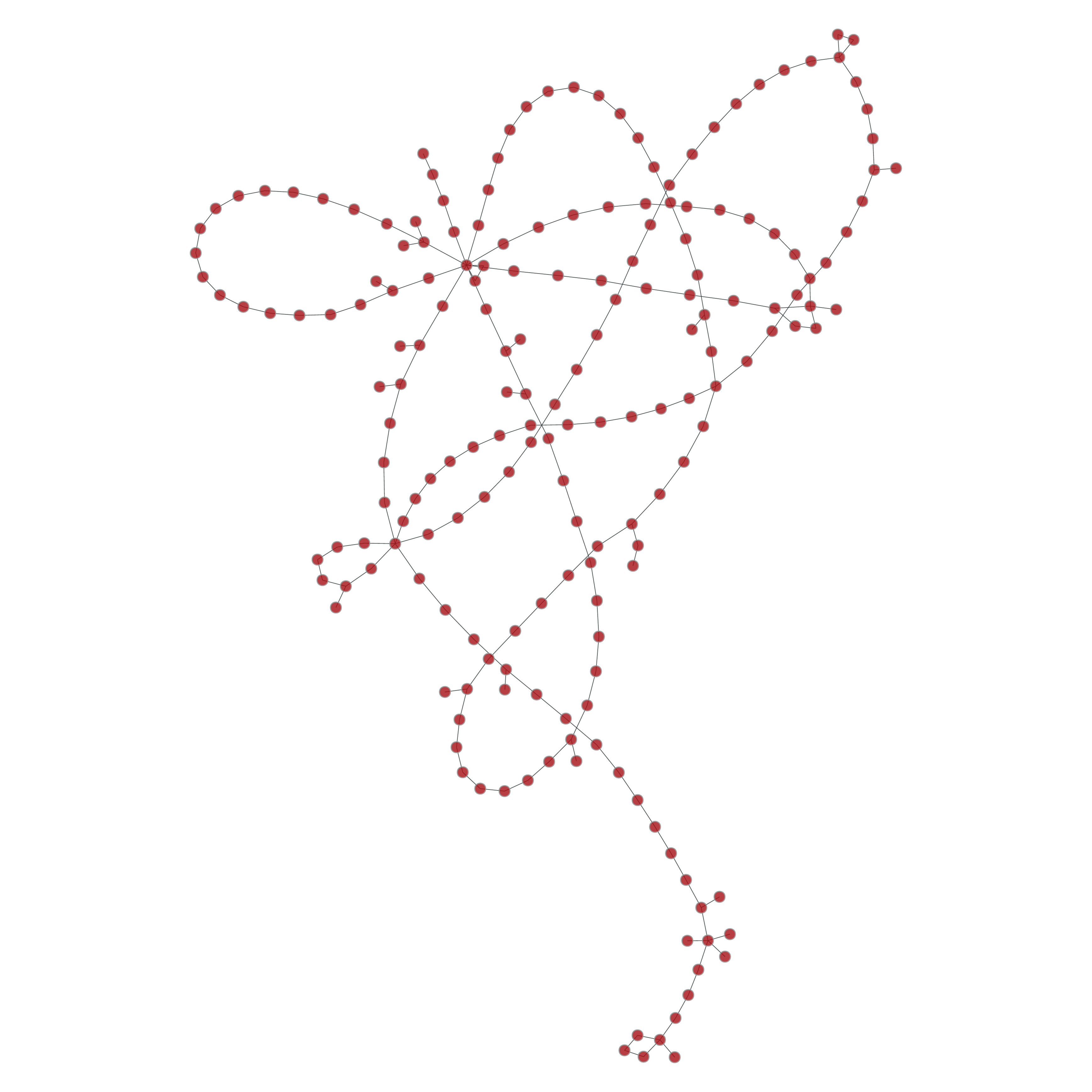}
		\caption{RW}
		\label{fig:graphrwfb15k-237}
	\end{subfigure}
	\begin{subfigure}{.2\textwidth}
		\includegraphics[width=0.9\linewidth]{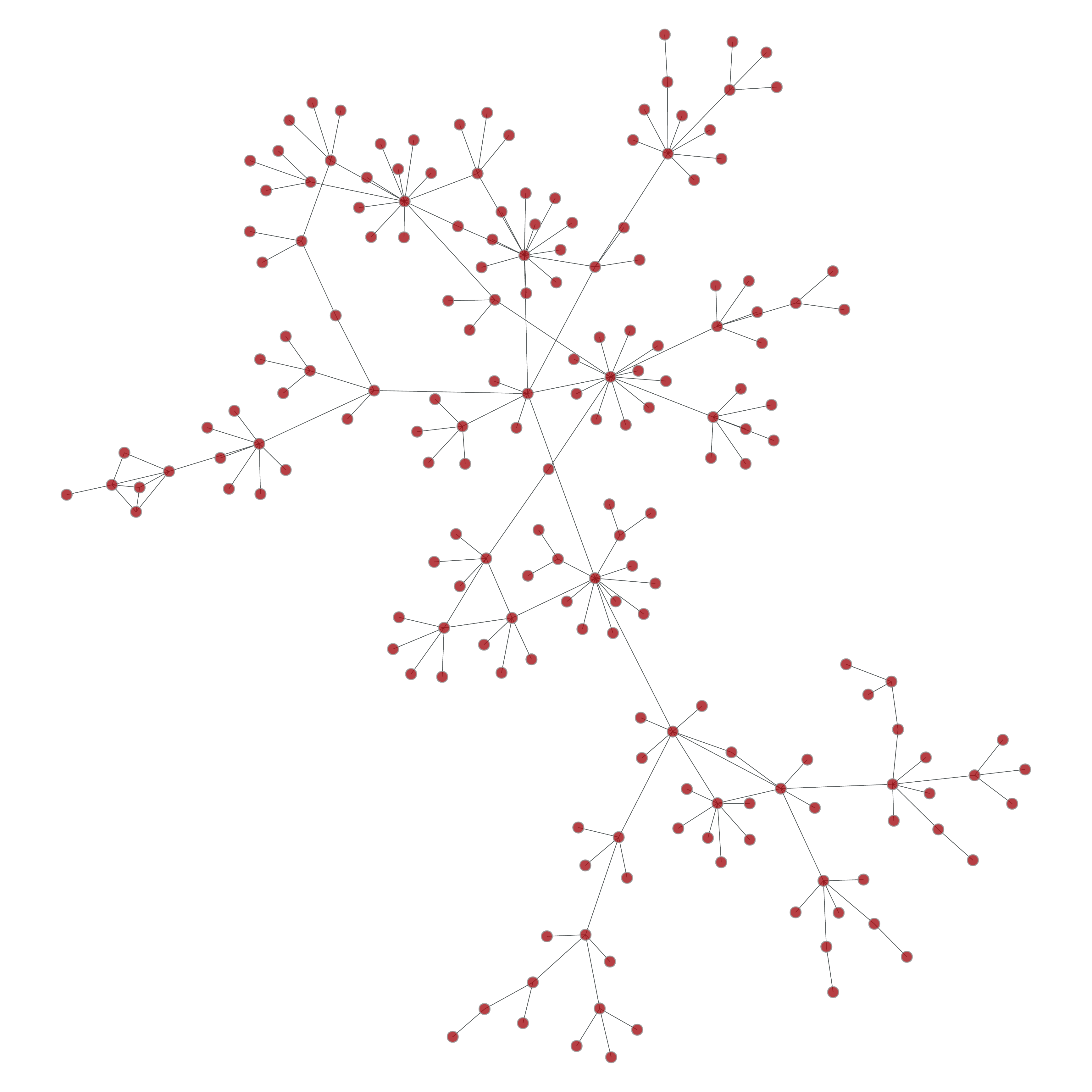}
		\caption{RWR}
		\label{fig:graphrwrfb15k-237}
	\end{subfigure}
	\begin{subfigure}{.2\textwidth}
		\includegraphics[width=0.9\linewidth]{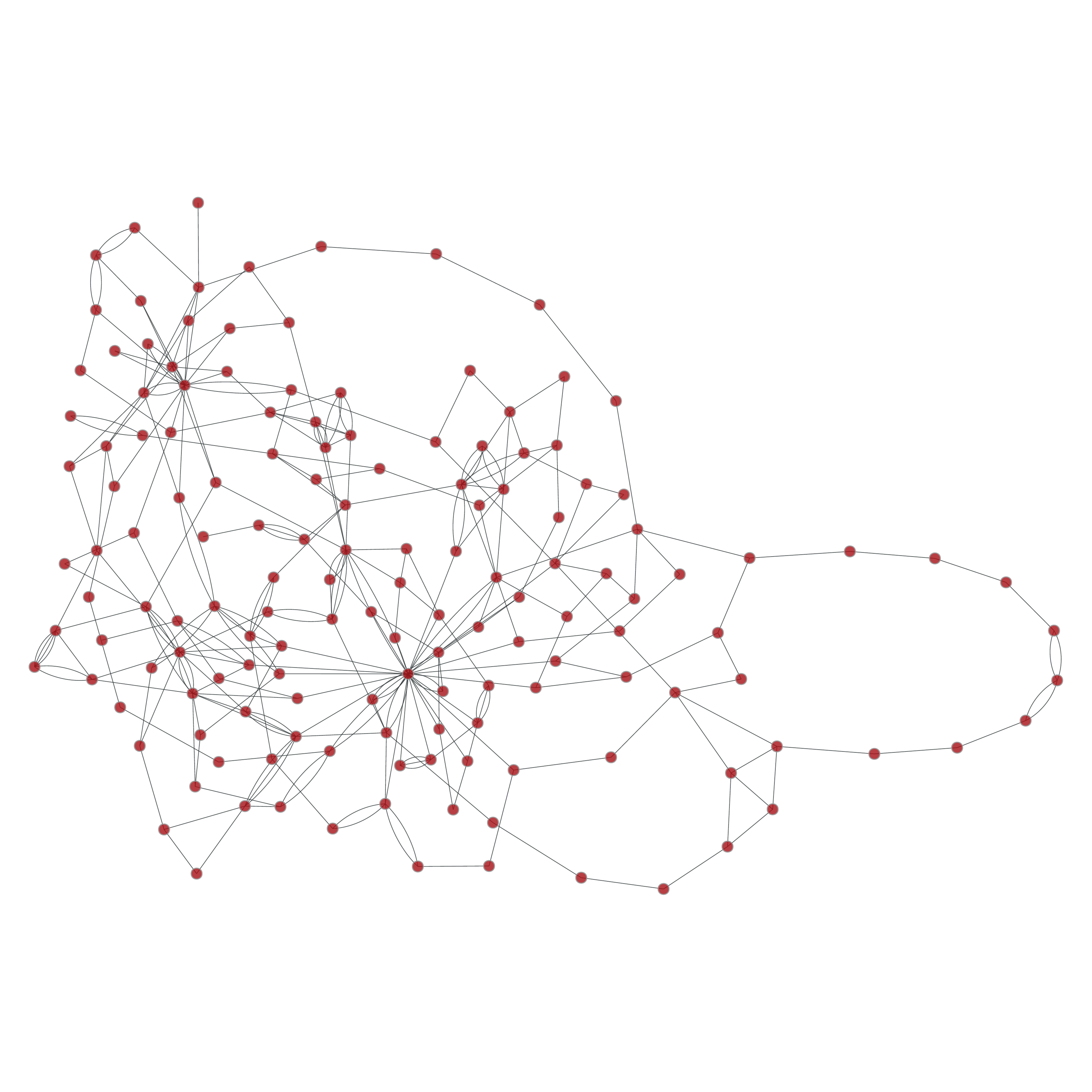}
		\caption{RWISG}
		\label{fig:graphrwisgfb15k-237}
	\end{subfigure}
	\begin{subfigure}{.2\textwidth}
		\includegraphics[width=0.9\linewidth]{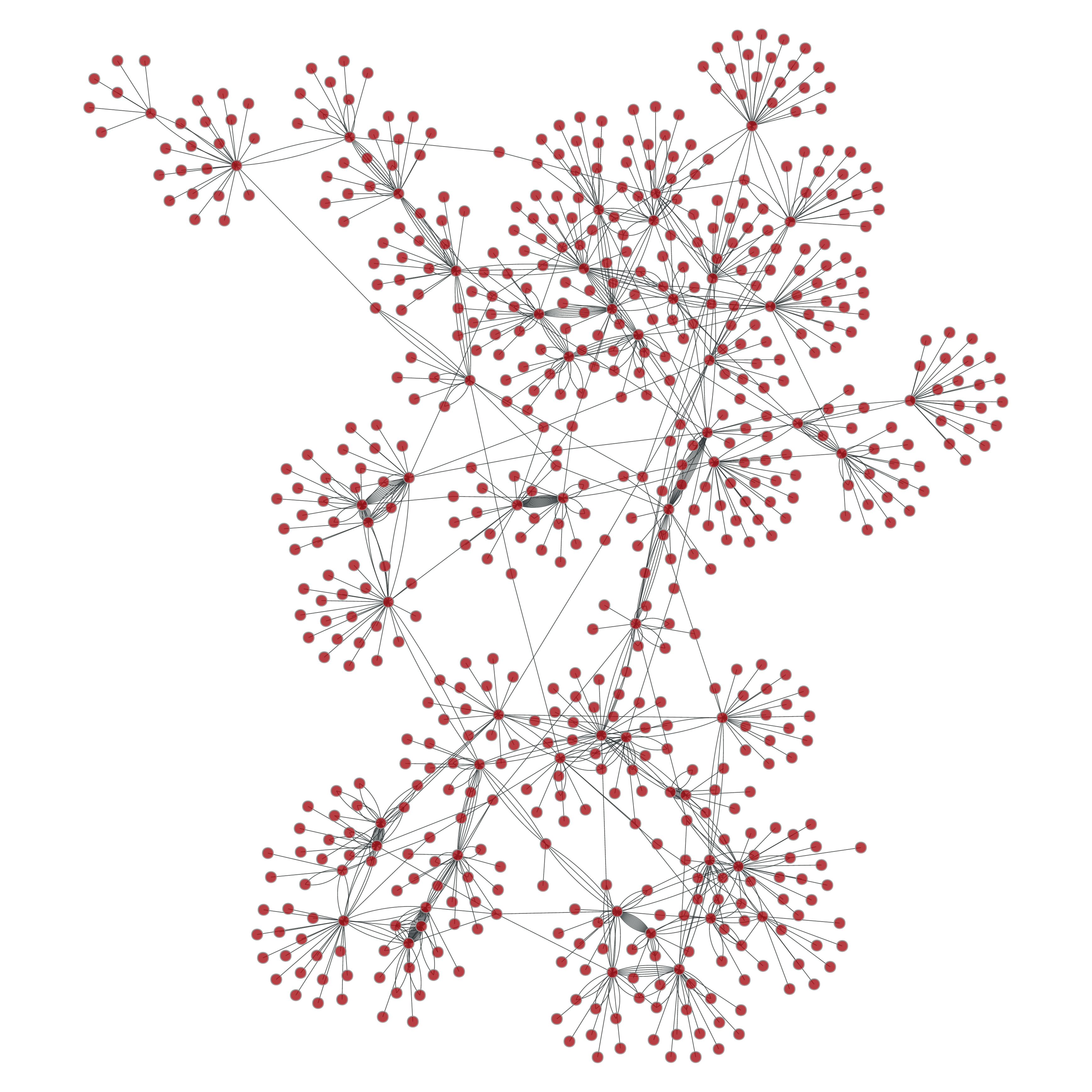}
		\caption{RWISG-N}
		\label{fig:graphrwrisgfb15k-237}
	\end{subfigure}
	\caption{Visualization for Minibatch-Subgraphs sampled using different algorithms [Samples are generated from DB100K knowledge graph]. The figures illustrate nature of real subgraphs constructed using minibatch of triples using different minibatch sampling algorithms proposed in this paper. The standard minibatch sampling algorithm followed in literature is shown as \textit{Simply Random} and samples mostly independent triples from the graph. The next four figures show the structure of the minibatch subgraphs obtained using various graph sampling approaches. More properties of these minibatch subgraphs are discussed in Section \ref{sec:minib-and-kg}}
\end{figure*}

\section{Introduction}
\label{sec:intro}
Knowledge Graph (KG) is a structured way of storing information in terms of various concepts and relations between those concepts. These concepts are referred to as entities and are represented as the vertices in a Knowledge Graph. The directed edges in this graph are the relations between two such connected concepts or entities. These edges are labeled by a relation type, which denotes the specific nature of relation between the two connected entities. 

Knowledge Graphs are built either using automatic information extraction tools like OpenIE~\cite{openie1,openie-mausam} or NELL~\cite{nell}, or from expert annotations like in FreeBase~\cite{freebase}, WordNet
~\cite{wordnet} or DBPedia~\cite{dbpedia} (source: Wikipedia infobox and other structured information). But because of reasons like, imperfections in the heuristics of IE based methods or time and investment required for expert annotations, or because of dynamic and temporal nature of certain relations, these Knowledge Graphs are not complete, i.e., they do not contain all the valid facts (represented using triplets of entity pairs and a relations) about the concepts that are present in these knowledge graphs. Knowledge Graph Completion is a task designed to tackle this issue of incompleteness. In this task, researchers build models that can assign some score or probabilities to all the possible missing facts such that the valid facts (including the missing ones) are assigned higher score or probability. This task is made possible by making use of the patterns in connectivity of different entities in the knowledge graph. For example, by knowing the home town of an athlete and which country that town is located in, with high probability we can tell which country this athlete represents. That is, although some facts are missing from these KGs, facts that are present in the knowledge graph contribute towards calculating the plausibility of missing ones.

Knowledge Graph Embedding~\cite{transe} is a powerful technique for learning a distributed representation for the entities and relations in a KG. In such a representation, the fixed dimensional entity vectors can encode information about its neighborhood, what types of relations does it have that connect it to the rest of the graph, its categorical information, etc. 
The most popular method for knowledge graph completion is by means of learning a knowledge graph embedding for this task itself. The model is trained to predict the valid facts in a KG, in terms of the distributed representations learned for the entities and the relations.
In a usual training regime, KGC models are trained by optimizing margin loss or cross-entropy loss using minibatch gradient descent. Under this setting, each minibatch consists of randomly sampled positive fact triplets and some artificially generated negative fact triplets. We will show that no matter the size of these randomly selected minibatches, majority of the entities within a minibatch occur only in one triplet. This means that the subgraphs formed by these minibatches are essentially very sparse graphs (see Figure \ref{fig:graphsrfb15k-237}). 
%Also, from these figures it is clear that the sparsity of the FB15k-237 knowledge graph is quite less than WN18RR, but the minibatches that are used for training are still really sparse. 
Hence, we hypothesize that training with minibatches consisting of randomly sampled fact triplets is not the best policy. In these minibatches, each entity occurs mostly atmost once, causing the updates to be biased only towards a single fact triplet and thus leading to a unstable updates \cite{nemirovski2009sgd}.

% \usepackage{booktabs}
% \begin{table*}[]
% 	\centering
% 	\begin{tabular}{@{}llll@{}}
% 		\toprule
% 		\textbf{Model(M)} & \textbf{Scoring Function,} $\phi_M(t)$ & & \\ \midrule
% 		\textbf{TransE} \cite{transe} & $\|e_s+r-e_o\|_2$ &  &  \\
% 		\textbf{DistMult} \cite{distmult} & $e_s^TW_re_o$ &  &  \\
% 		\textbf{ComplEX} \cite{complex} &  &  &  \\
% 		\textbf{RotatE} \cite{rotate} &  &  &  \\ \midrule
% 		\textbf{Loss Function} & $L = -\frac{1}{2} \left[\log\sigma (\phi (t_+)) + \log\sigma (-\phi (t_-))\right]$ &  &  \\ \bottomrule
% 	\end{tabular}
% 	\caption{}
% 	\label{tab:model-props}
% \end{table*}

In this paper, we first empirically show that randomly sampled minibatches have a very poor degree distribution which leads to high variance of loss gradients and sub-optimal convergence of the model. Next we show that, although in small scale KGs, like FB15k-237 and WN18RR, using a large minibatch size can mitigate these problems, the problem becomes more pronounced in large scale KGs (e.g. DB100K) and having large enough batch size becomes a computational limitation. Next, we propose a random walk based minibatch selection method which can be applied for any model (that relies on MBSGD) and dataset of any scale. Finally, we share the results for thorough experiments with three different models and with three knowledge graphs of different scale and structure. Our model provides large improvements for the large scale knowledge graph DB100K. 

%In this paper, we first show that including neighboring triplets of the entities in a minibatch helps reduce the sparsity in the minibatch. Based on this observation we formulate a new generalized loss function that takes into this nature of minibatches used for training. We further show that using this new loss function better performance can be achieved for many different types of KGC models.

\section{Background}
\label{sec:background}
A knowledge graph can be represented as a triple $G=\{E,R,T\}$. Here, $E$ and $R$ are the sets of all entities and relations, and $T$ is the set of all fact triplets in the knowledge graph. Individual facts are triplets of the form $t = (s, r, o)$ for some $t \in T$, $s,o\in E$ and $r \in R$. Note that such fact triples are directional, i.e., $s$ has a relation $r$ to $o$ but this is not necessarily true the other way.

In a knowledge graph embedding model, for the completion task, each entity $s\in E$ and relation $r\in R$ has a vector representations $e_s$ and $w_r$, respectively. For the completion task, the decoder consists of a scoring metric $\phi$ for the triplets $t \in T$. In this section, we will first look into the loss functions used by different models for the knowledge graph completion task.

% The generic loss function,

% \begin{align}
%     L &= -\sum_{\substack{(h,r,t) \in T_m \\ (h',r,t') \notin T}} f(d(h,r,t), d(h',r,t'))
% \end{align}

\noindent \textbf{TransE}: In TransE \cite{transe}, a relation between two entities is interpreted as a relation specific translation from the subject to the object entity, in a low dimensional embedding space, where each entity is uniquely represented. The score function used in TransE is, 

\begin{align}
    \phi_{TransE} = -\|e_s+w_r-e_o\|_2
\end{align}

\noindent \textbf{Distmult}: Proposed by \cite{distmult}, in Distmult model the score of a triplet $(s,r,o)$ is measured as a weighted bilinear product between the head and tail entity embedding. The diagonal weight matrix is specified by the relation $r$. 

\begin{align}
    \phi_{Distmult} = e_s^T W_r e_o \\
    W_r = diag(w_r) \nonumber
\end{align}

where $W_r$ is a diagonal matrix with $w_r$ as the diagonal. Dismult is trained with margin based ranking loss function.

\noindent \textbf{ComplEx}: Unlike TransE and Distmult, the ComplEX model, proposed by \cite{complex}, uses complex vectors for representing entities and relations. The score function for ComplEX is,

\begin{align}
    \phi_{ComplEX} &= Re(\langle w_r, e_s, \Bar{e_o} \rangle) \\
    w_r, e_s, e_o  &\in C^K \nonumber
\end{align}

\noindent \textbf{RotatE}: Proposed by \cite{rotate}, the RotatE model uses relation specific rotations in complex domain to link between the subject and the object entities. The expression for RotatE's loss function is,

\begin{align}
    \phi_{RotatE} = -\|e_s*w_r - e_o\|_2
\end{align}

where, $e_s, w_r, e_o \in \mathbb{C}^K$.

% \subsection{KBAT}
% % Encoder is Multi relation GAT and decoder is ConvE
% Unlike the previous models the encoder in KBAT\cite{kbat} is not a fixed embedding matrix. Its encoder is based on Graph Attention Network (GAT) originally proposed for homogeneous graphs \cite{gat}. In GAT, after every layer the entities gets a new embedding by attending over the embedding (in previous layer) of its neighbors. \cite{kbat} extends this idea to knowledge graph, which is heterogeneous in nature, by considering the relation types also for computing the attention over the neighborhood entities. The following loss function is used for training this encoder,

% \begin{align}
%     L(\Omega) = \sum_{t_{ij} \in S} \sum_{t'_{ij} \in S'} \max \{d_{t'_{ij}} - d_{t_{ij}} + \gamma, 0\}
% \end{align}

% KBAT reuses the decoder from ConvE model \cite{convE} for KGC and trains it using the following loss function.
% Decoder Loss:

% \begin{align}
%     L(\Omega) = \sum_{t^k_{ij} \in S \cup S'} log(1+ \exp (l_{ij}^k f(t_{ij}^k)) ) + \frac{\lambda}{2}\|W\|^2
% \end{align}

In the following section, we will explain the optimization algorithm used for training KGC models in details.

\begin{figure*}[h]
	\begin{minipage}[t]{.49\textwidth}
		\centering
		\includegraphics[width=\textwidth]{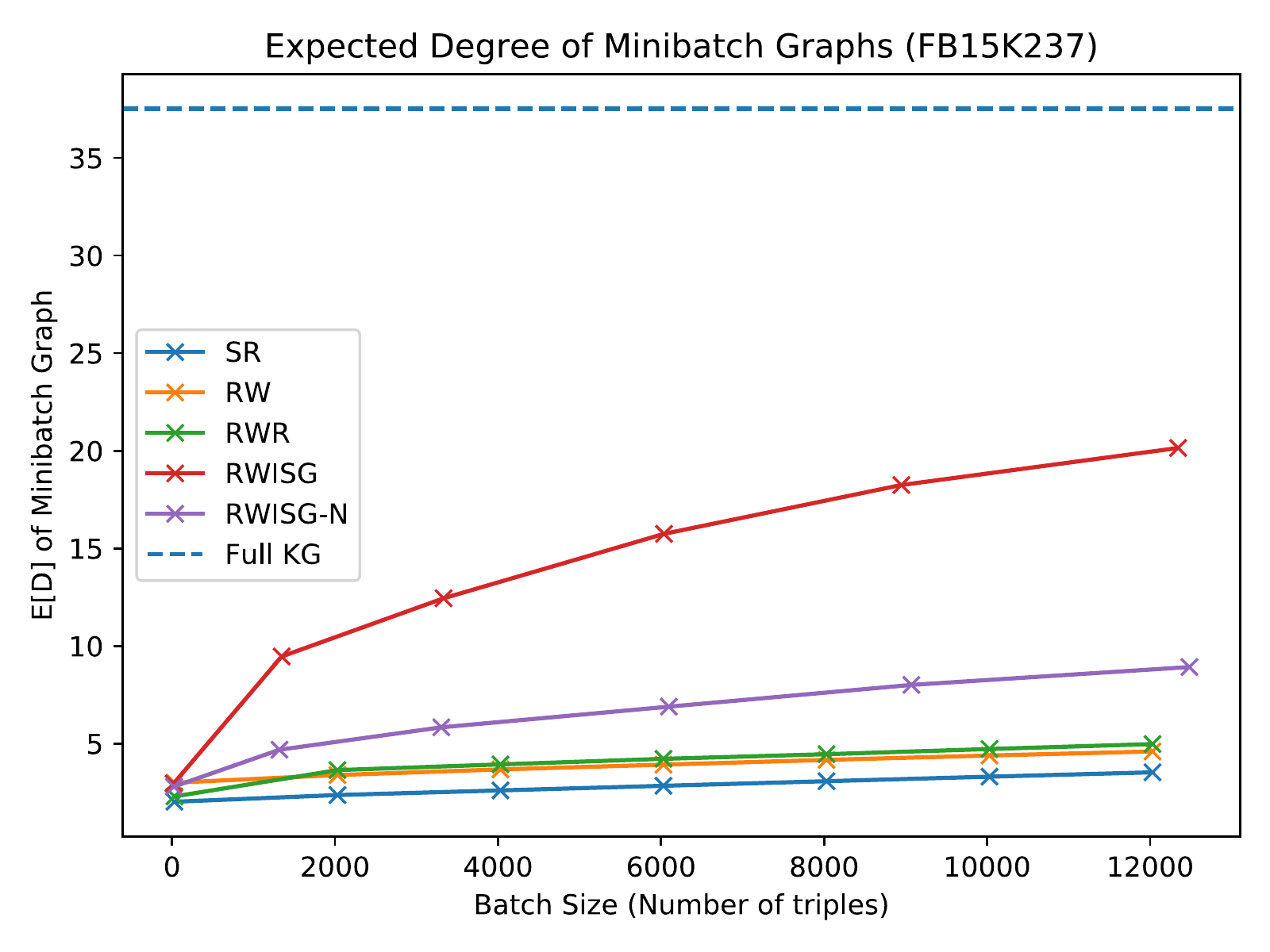}
		% \subcaption{Image 1.}\label{fig:1}
	\end{minipage}
	\hfill
	\begin{minipage}[t]{.49\textwidth}
		\centering
		\includegraphics[width=\textwidth]{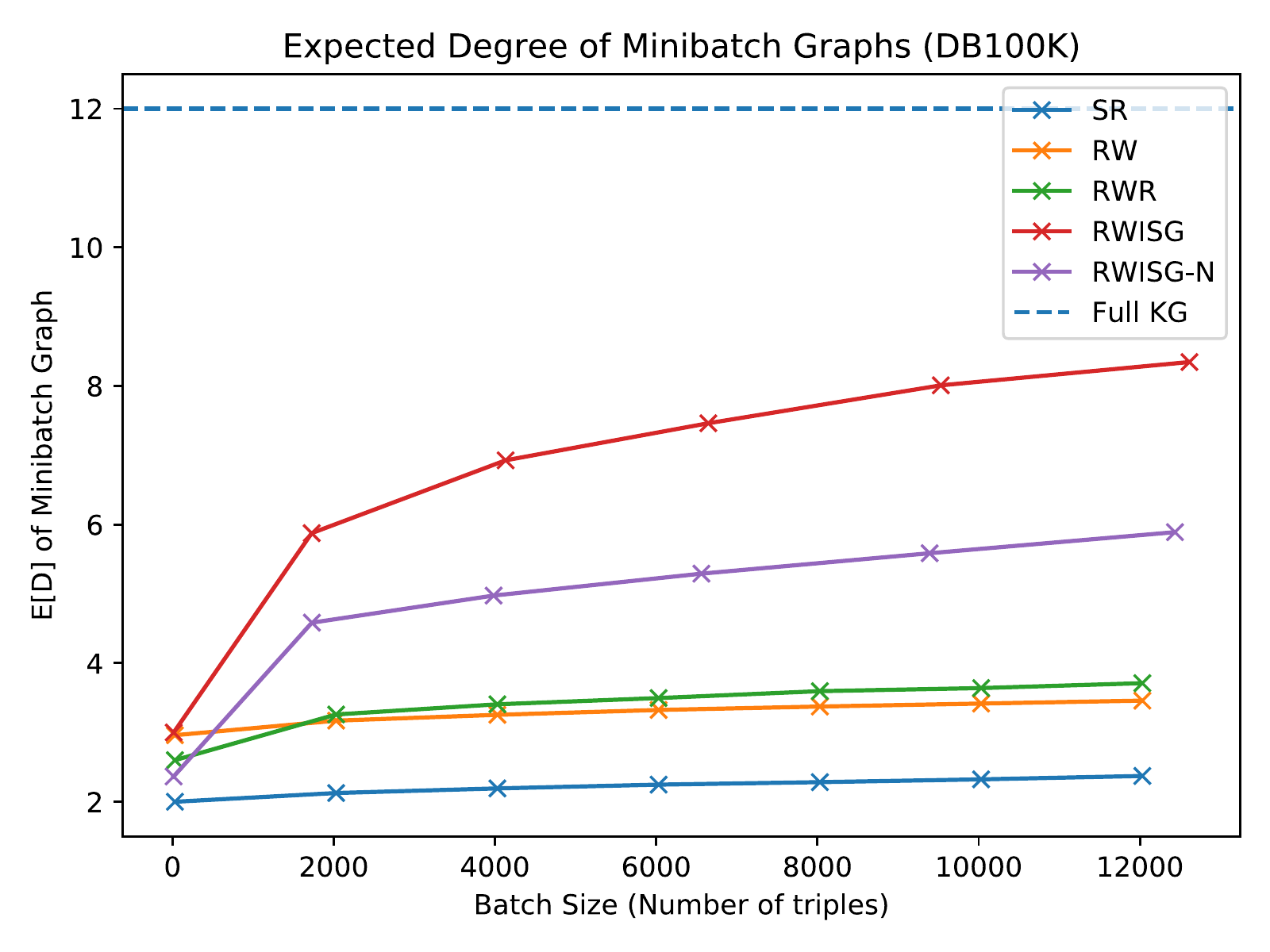}
		% \subcaption{Image 2.}\label{fig:2}
	\end{minipage}
	\caption{Expected degree of vertices in minibatch for different sampling algorithms. The low values for $E[D]$ in case of Simply Random (SR) indicates that each entity in these minibatches co-occurs only with few triplets.}
	\label{fig:ed-vs-bs}
\end{figure*}

\begin{figure*}[h]
	\begin{minipage}[t]{.49\textwidth}
		\centering
		\includegraphics[width=\textwidth]{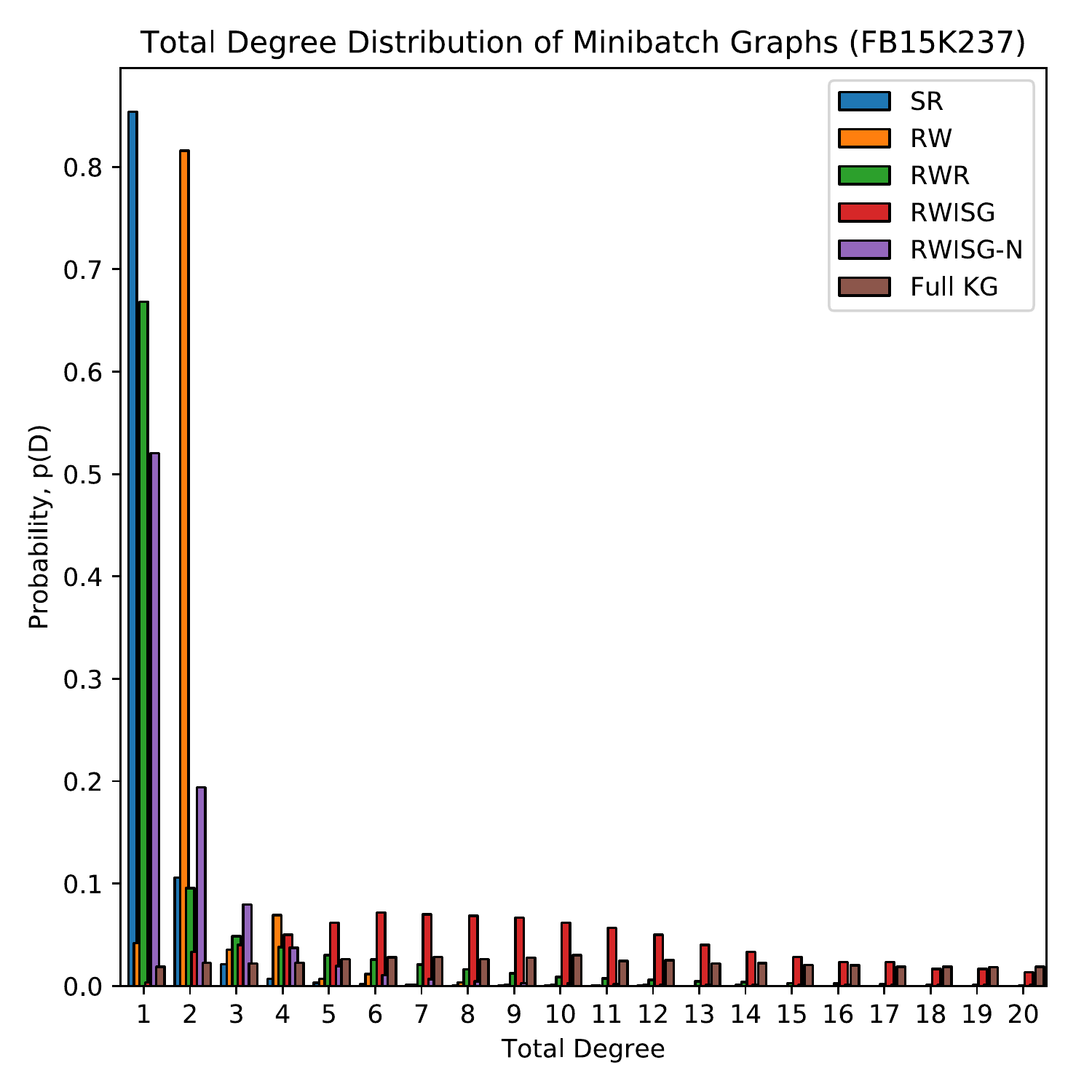}
		% \subcaption{Image 1.}\label{fig:1}
	\end{minipage}
	\hfill
	\begin{minipage}[t]{.49\textwidth}
		\centering
		\includegraphics[width=\textwidth]{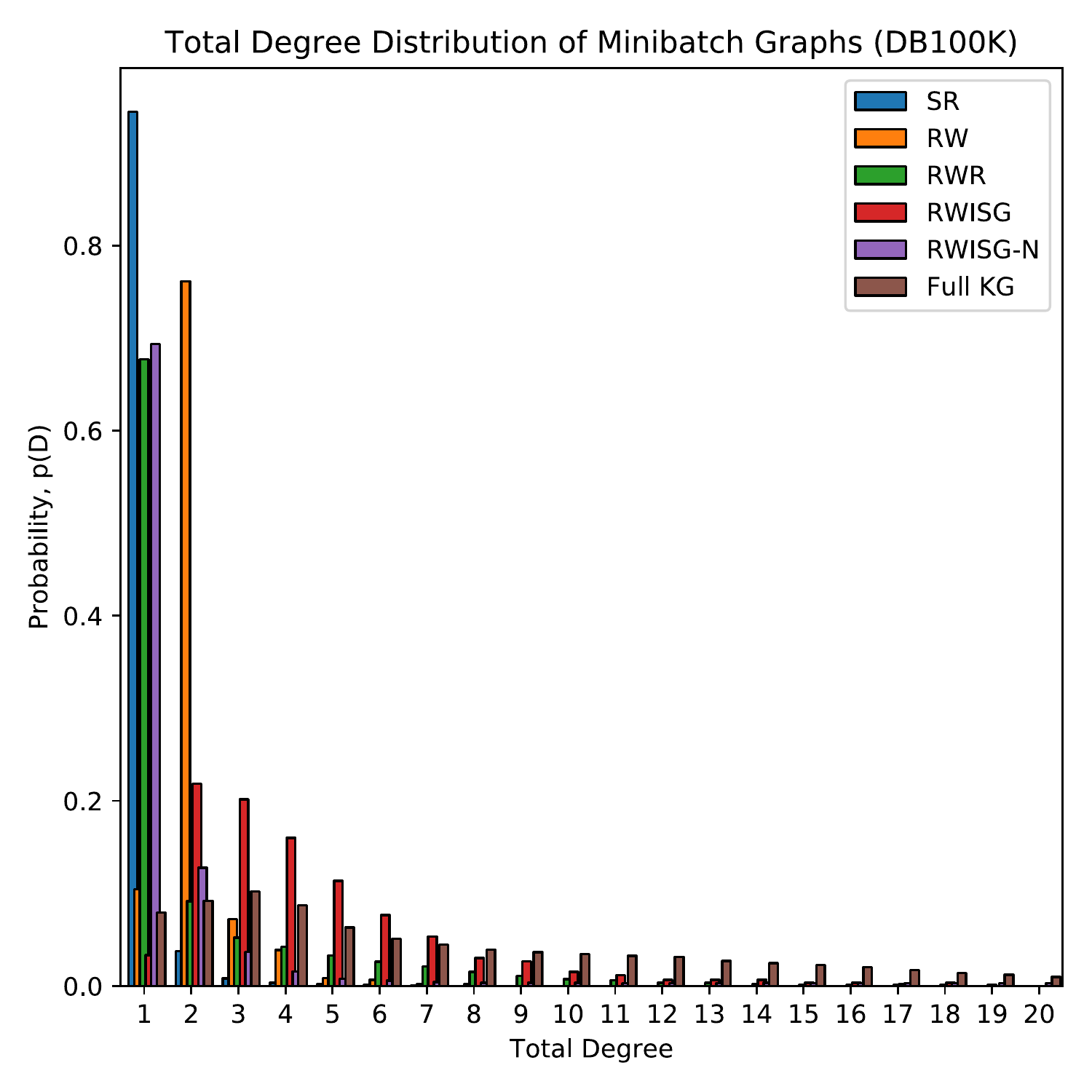}
		% \subcaption{Image 2.}\label{fig:2}
	\end{minipage}
	\caption{Comparison of total degree distribution of minibatches from different sampling algorithms [Tail of the distributions beyond $D=20$ is truncated for better clarity]. Total degree $d=0$ is omitted as any entity occurs as part of atleast one triple.}
	\label{fig:degree-distributions}
\end{figure*}

\section{Minibatch Sampling and Knowledge Graph Completion}
\label{sec:minib-and-kg}
In recent deep learning literature, \textit{Minibatch Stochastic Gradient Descent} \textbf{(MBSGD)} is highly prevalent, mainly because of reasons \textit{borrowed} from convex optimization literature and its ability towards stabilizing learning. In case of convex loss functions, when (full) batch gradient descent is not a realizable option, using MBSGD \cite{dekel2012mbsgd} in place of SGD \cite{nemirovski2009sgd} provides better convergence and rate of convergence.

Similarly, all the models described in previous section (and many others) are trained using MBSGD\footnote{Now-a-days more sophisticated optimizers like Adagrad, ADAM, etc. are used in place of SGD. Minibatches are used with these methods as well.} optimization algorithm. It involves, first, sampling a small number of positive samples of fact triples from the knowledge graph and then artificially creating a set of negative fact triples by corrupting the positive ones. These positive and negative samples together form a minibatch. Loss is then computed for a minibatch based on the scores assigned by the model for the samples within minibatch. Model parameters are then updated according to the gradient of the loss function calculated on this minibatch of samples. 

In deep learning models, where loss functions are very often non-convex, MBSGD is still used to avoid the computational barrier of full-batch gradient descent (for large datasets) and unstable gradients in stochastic gradient descent. By taking a minibatch of $b$ training samples, MBSGD decreases the variance in the estimate of gradient by a factor of $1/b$, compared to SGD, for model parameters which is used for calculating the parameter updates \cite{nemirovski2009sgd,dekel2012mbsgd}.

Now, if we consider the objective functions used for training the knowledge graph completion models described above, those can be written in the following generalized form,

\begin{align}
J(\theta) = \sum_{t \in G} L_\theta(t) \label{eqn:vanilla-loss}
\end{align}

where $G$ denotes the knowledge graph, $t$ denotes a triple in $G$ and $\theta=\{\mathcal{E}, \mathcal{W}_R\}$ is the set of parameters. $\mathcal{E}$ is the set of entity embedding vectors and $\mathcal{W}_R$ is the set of relation embedding vectors. The loss function $L_\theta$ is usually designed to maximize the score $\phi(t)$ for a correct triple $t$ (see Section \ref{sec:background}) against the incorrect ones. As a knowledge graph only stores an incomplete set of positive facts and no negative facts, the above loss is calculated using artificially generated negative samples. For instance, the objective function for soft-margin or log-sigmoid loss is

\begin{align}
	J(\theta) = \sum_{t \in G} -\frac{1}{2} \big[&\log\sigma(\phi(t) - \gamma) \nonumber \\ 
	&+ \mathbb{E}_{t_-\in T'_t}[\log\sigma(\gamma - \phi(t_-))]\big]
\end{align}

where, $T'_t$ denotes a subset of negative samples created by corrupting the triplet $t = (s,r,o)$.

\begin{align*}
    T'_{t=(s,r,o)} \subset \{(s',r,o)|s' \not= s \land (s',r,o) \not\in G\} \cup\\
    \{(s,r,o')|o' \not= o \land (s,r,o') \not\in G\}
\end{align*}

The gradient of the objective function with respect to any entity embedding vector\footnote{The only parameters involved in a KGC model are entity and relation embedding vectors.}, say $e_i$, is estimated using a minibatch of triples $T_m$. Note that, this gradient $\nabla_{e_i} J(\theta)$ will only have contributions from those triples in the minibatch that involve the entity $e_i$.

\begin{align}
	\nabla_{e_i} J(\theta) = \nabla_{e_i} \sum_{t \in T_m} L(t) = \sum_{\substack{t \in T_m \\ e_i \in T_m \cup T'_t}} \nabla_{e_i} L(t)
\end{align}

Since corruption is done using entities chosen randomly, the contribution towards gradient from $T'_t$ will occur very sparsely. Hence, the minibatch gradient for a single entity embedding vector is calculated using the number of triples containing the entity $e_i$ (instead of $b$ triples, which is the size of minibatch). Because of this reason, the improvement in the variance of updates when using MBSGD for KGC models is only of the order $\mathcal{O}(E[D]^{-1})$, $E[D]$ being the expected total degree of entities in a minibatch. 

\section{Proposed Method}

By increasing the minibatch size in \textit{Simply Random}\footnote{We refer to the standard method of selecting $b$ random triples from the training set as \textit{Simply Random} selection method.} selection method, $E[D]$ cannot be increased. Hence, in order to get denser minibatch subgraphs, we propose the use of sampling methods from the stochastic graph sampling literature \cite{leskovec2006sampling}. In this context we consider the following graph sampling algorithms.

\subsection{Graph Sampling Methods}
\label{sec:sampling-algo}
\paragraph{Simply Random (SR)} By \textit{Simply Random} minibatch selection, we refer to the standard policy of randomly selecting $b$ triples in the knowledge graph completion literature.

\paragraph{Random Walk (RW)} As the name suggests, this graph sampling method is based on random exploration of a graph. Sampling starts with first selecting a random initial vertex. After that, at every step the sampler moves to one randomly selected neighbor of the current vertex \cite{leskovec2006sampling}. As shown in Figure \ref{fig:graphrwfb15k-237}, samples from a large knowledge graph usually take chain like structure (with a few junctions and open ends) and almost every vertex participates in exactly two triples. Because of this, $E[D] \approx 2$ for sample size (same as minibatch size in our case) $b \ll |G|$. The trend of $E[D]$ with minibatch size is depicted in Figure \ref{fig:ed-vs-bs}.

\paragraph{Random Walk with Restart (RWR)} To avoid this chaining effect in the sample and obtain a more dense subgraph, a well known method is Random Walk with Restart where at each stage of sampling, the sampler jumps back to (restarts from) some previously selected node (usually fixed to the starting node). 

\paragraph{RW with Induced Subgraph Sampling (RWISG)} To mitigate the problems faced in RW and RWR, \cite{lu2012sampling}\footnote{In the original publication, authors have referred to this algorithm as Neighborhood Reservoir Sampling.} proposed using the Induced Subgraph (ISG) as the sample. In this sampling algorithm, after selecting the vertices through the RW method, the subgraph induced by these vertices is taken as the final minibath sample.

% \paragraph{RWR with Induced Subgraph Sampling (RWRISG)} This is a variant of the RWISG algorithm using RWR instead of RW. This had very similar $E[D]$ and model performance to that of RWISG only. 

\paragraph{RWISG-N} We formulated this method as an intermediate between RW and RWISG. RW greedily adds any neighbor of the current node whereas at completion RWISG returns the induced subgraph of the vertices selected. RWISG-N returns a subgraph which is union of the induced subgraph and a set of randomly selected neighbors of the vertices. Its expected degree is between RW and RWISG.

% \paragraph{RWR with Induced Subgraph Sampling (RWRISG)}

\subsection{Degree Distributions of Minibatch Subgraphs}
We compare the degree distribution of minibatch subgraphs (see Figure \ref{fig:degree-distributions}) sampled using the different algorithms as described in Section \ref{sec:sampling-algo}. This gives us an idea about how dense the minibatch subgraphs are. Also for ease of comparison we have included the degree distribution for the whole knowledge graph as well. For SR sampling algorithm, more than 80\% of the entities have degree $1$ in the minibatch. For RW, we see a peak in the distribution for degree $2$ because of its chain like structure. But when we consider RWR, we again see a surge in probability of degree $1$ because of the dangling entities near the outer perimeter of the subgraph (see Figure \ref{fig:graphrwrfb15k-237}). This problem is solved upto a large extent by replacing the set of triples selected by RW or RWR with the subgraph induced by the entities. This can be seen for the final two algorithms RWISG and RWISG-N. 

We have used multiple minibatch samples to calculate the degree distributions as average of the degree distribution of the individual minibatch subgraphs.
\begin{align}
 P_D^{(i)}(d) &= \frac{|{v:v \in G_s^{(i)}.V, D(v)=d}|}{|G_s^{(i)}.V|} \\
 P_D(d) &= \frac{1}{N} \sum P_D^{(i)}(d)
\end{align}
Also, since $\sum_{d=0}^{\infty} P_D(d) = 1$, $P_D(d)$ is a probability distribution.

The distributions in Figure \ref{fig:degree-distributions} are drawn for a fixed batch size. These distributions would also change as the minibatch sample size $b$ takes different values (possibly upto $|G.E|$). To understand this effect we calculate expected total degree $E[D]$ of the minibatch subgraphs for each sampling algorithm. Also, since each draw of minibatch sample can be noisy, we first obtain an average histogram using several draws and then calculate $E[D]$.

\begin{align}
 E[D] = \sum_{d=1}^{d_{max}} P_D(d)*d
\end{align}

Figure \ref{fig:ed-vs-bs} shows how expected degree of minibatch subgraphs varies with minibatch size for different sampling algorithms. These were empirically calculated from a large number of minibatch samples. We observe very similar trends and RWISG gives much better $E[D]$ that the other algorithms.

%\begin{example}[How to write an example]
%Examples should be written using the example environment defined in this template.
%\end{example}

%\begin{theorem}[A titled theorem]
%Using RWR-Minibatch selection algorithm with same batch size $b$ decreases variance of %minibatch gradient estimate.
%\end{theorem}

%\begin{proof}
%Small proof goes here.
%\end{proof}

% \begin{algorithm}[tb]
% \caption{Proposed Minibatch Selection Algorithm}
% \label{alg:algorithm}
% \textbf{Input}: Undirected Graph(G)\\
% \textbf{Parameter}: Minibatch Size (B)\\
% \textbf{Output}: Minibatch Subgraph $G_s$
% \begin{algorithmic}[1] %[1] enables line numbers
% \STATE Let $l=0$
% \WHILE{$l \le B$}

% \IF {$l$ not updated for 5 steps}
% \STATE $v = sampleVertex(G)$
% \ENDIF

% \STATE nxt = sampleNext(G, v)

% \IF {$(v,nxt)$ not in $G_s$}
%     \STATE $G_s \gets G_s\cup (v,nxt)$
% \ENDIF

% \STATE $v \gets nxt$
% \STATE $l \gets G_s.size()$

% \ENDWHILE
% \end{algorithmic}
% \end{algorithm}

\section{Experiments and Results}

\begin{table}[]
\centering
\begin{tabular}{L{2.2cm}R{1.7cm}R{1.5cm}R{1.4cm}}
\toprule
\textbf{Dataset} & \textbf{FB15k-237} & \textbf{WN18RR} & \textbf{DB100K} \\ \midrule
\textit{\#Entities} & 14,541 & 40,943 & 99,604 \\ 
\textit{\#Relations} & 237 & 11 & 470 \\ 
\textit{\#Train} & 272,115 & 86,835 & 597,572 \\ 
\textit{\#Validation} & 17,535 & 3,034 & 50,000 \\ 
\textit{\#Test} & 20,466 & 3,134 & 50,000 \\ 
\textit{Avg. Degree} & 37.4 & 4.2 & 12.0 \\ 
\textit{Median Degree} & 22 & 3 & 7.0 \\ \bottomrule
\end{tabular}
\caption{Properties of FB15k-237, WN18RR and DBPedia100K Knowledge Graphs. Average degree is calculated using total degree of vertex, which is the sum of in-degree and out-degree of a vertex.}
\label{tab:kgstats}
\end{table}

\begin{table*}[h]
\centering
\begin{tabular}{llrrrrr}
\toprule
\textbf{Model} & \textbf{Minibatch Selection Method} & \textbf{MRR} & \textbf{MR} & \textbf{Hits@1} & \textbf{Hits@3} & \textbf{Hits@10} \\
\toprule
\multicolumn{7}{c}{\textbf{Dataset: DB100k}} \\
% \midrule
% \textbf{TransE} & \textit{Vanilla} & 0.111 & - & 0.016 & 0.164 & 0.270 \\
% \textbf{DistMult} & \textit{Vanilla} & 0.233 & - & 0.115 & 0.301 & 0.448 \\
\midrule
\multirow{3}{*}{\textbf{ComplEX}} &  \textit{Vanilla} & 0.232 & 1740 & 0.150 & 0.265 & 0.385 \\
 & \textit{RWISG} & 0.219 & 2203 & 0.143 & 0.252 & 0.362 \\
 & \textit{RWISG-N} & \textbf{0.254} & \textbf{1171} & \textbf{0.168} & \textbf{0.292} & \textbf{0.411} \\
 \midrule
 &  \textit{ComplEx-NNE+AER} & 0.306 & - & 0.244 & 0.334 & 0.418 \\
\midrule
\multirow{3}{*}{\textbf{RotatE}} & \textit{Vanilla} & 0.296 & 2614 & 0.169 & 0.377 & 0.514 \\
 & \textit{RWISG} & 0.347 & \textbf{844} & 0.209 & 0.439 & 0.584 \\
 & \textit{RWISG-N} & \textbf{0.396} & 937 & \textbf{0.275} & \textbf{0.474} & \textbf{0.604} \\
\toprule
\multicolumn{7}{c}{\textbf{Dataset: FB15k-237}} \\
\midrule
\multirow{3}{*}{\textbf{TransE}} & \textit{Vanilla} & 0.292 & \textbf{180} & 0.198 & 0.327 & \textbf{0.48} \\
 & \textit{RWISG} & \textbf{0.296} & 201 & \textbf{0.204} & \textbf{0.330} & 0.478 \\
 & \textit{RWISG-N} & 0.294 & 185 & {0.202} & {0.328} & 0.475 \\
\midrule
\multirow{3}{*}{\textbf{DistMult}} & \textit{Vanilla} & 0.241 & 254 & 0.155 & 0.263 & \textbf{0.419} \\
 & \textit{RWISG} & \textbf{0.249} & \textbf{242} & \textbf{0.175} & 0.269 & 0.397  \\
 & \textit{RWISG-N} & 0.250 & 231 & 0.172 & \textbf{0.273} & 0.403 \\
%  DistMult RWISG-NLoss 0.250,231,0.172,0.273,0.403
\midrule
\multirow{3}{*}{\textbf{RotatE}} & \textit{Vanilla} & 0.338 & \textbf{177} & 0.241 & 0.375 & 0.53 \\
 & \textit{RWISG} & 0.334 & 190 & 0.238 & 0.372 & 0.53  \\
 & \textit{RWISG-N} & \textbf{0.343} & 185 & \textbf{0.249} & \textbf{0.377} & \textbf{0.532} \\
\toprule
\multicolumn{7}{c}{\textbf{Dataset: WN18RR}} \\
% \midrule
% \multirow{3}{*}{\textbf{DistMult}} & \textit{Vanilla} & 0.43 & 5110 & 0.390 & 0.440 & 0.490 \\
%  & \textit{RWISG} & & & &  \\
%  & \textit{RWISG-N} & & & &  \\
\midrule
\multirow{3}{*}{\textbf{RotatE}} & \textit{Vanilla} & \textbf{0.476} & \textbf{3340} & 0.428 & {0.492} & 0.571 \\
 & \textit{RWISG} & 0.476 & 3396 & 0.428 & \textbf{0.494} & \textbf{0.572} \\
 & \textit{RWISG-N} & 0.474 & 4108 & \textbf{0.434} & 0.485 & 0.555  \\
% wn18 for RotateE RWISG 0.476, 3396, 0.428,0.494,0.572
\bottomrule
\end{tabular}
\caption{Performance comparison of various KGC models with different minibatch selection strategies. The baseline performance numbers are based on reproduction from the codes published by authors of the respective papers. Baseline results for DB100K are taken from \protect\cite{ding-2018-db100-cmplx} or by reproducing the experiments with OpenKE, whichever is better. ComplEx-NNE+AER is the best published model for DB100K. For the omitted models, the results for RWISG and RWISG-N were almost the same as with vanilla. The numbers in bold indicate the best performance in terms of a metric for every dataset-model pair.} %\cite{ding-2018-db100-cmplx}
\label{tab:results-all}
\end{table*}

% 0.473057,4122.785,0.431078,0.487875,0.555361,128680300,20000

%\paragraph{Modified MBSGD}
To use these proposed minibatch sampling algorithms during training, we replaced the minibatch in each training iteration with a sample drawn from these samplers. All the different algorithms in Section \ref{sec:sampling-algo} showed how these led to minibatch subgraphs with varying connectivity. %Among these, we report results for SR, RWISG and RWISG-N methods only as for the rest of the methods' performance was very close to either of the three above.

% TODO: 
% describe WN18 -> WN18RR and same of FB
% Add dataset stats. mainly details about neighborhoods -> try to exlain

We evaluate our proposed minibatch selection algorithm with different KGC models and several datasets of different scales and sparsity to fully understand its effectiveness. For the different models and dataset, we first obtain a benchmark result by running the models from two popular open-source repository, OpenKE \cite{han2018openke} and RotatE \cite{rotate}. These are mentioned in Table \ref{tab:results-all} as \textbf{Vanilla} models. Note that, Vanilla corresponds to using the SR sampler as the models used a random shuffle of the training dataset as the sequence of training minibatches. We have further modified each of the repositories to train any of the available models with the proposed minibatch selection algorithm, \textbf{RWISG} and \textbf{RWISG-N}. 

%We ran experiments only with Simply Random, RWISG and RWISG-N as the remaining two algorithms RW and RWR are comparatively slower. Following table shows a comparison between the complexity of the sampling methods.

\subsection{Datasets}
\paragraph{FB15k-237}
FreeBase is a large knowledge base consisting of RDF style fact triplets sourced from various structured content (e.g. structured data in Wikipedia submitted by users) on the web. It contains 1.9 billion triplets. For the purpose of evaluation of KGC models, \cite{transe} release a smaller version with 15k entities called FB15k, which has been used since as a standard for evaluating KGC models. Later, \cite{fb15k237-fb-features} released a cleaned version of FB15k called FB15k-237, since the original version contained data leaks in the test set due to inverse relations. 

\paragraph{WN18RR}
WN18RR is a subset of WordNet lexical database \cite{convE}. Originally proposed as WN18, this graph is hierarchical in nature with relations like hypernyms, meronymy, etc. But similar to FB15k, this knowledge graph also had the problem of inverse or duplicate links. \cite{convE} resolves those problems in WN188RR, by removing redundant inverse relations from the dataset. 

\paragraph{DB100K}
DB100K, released by \cite{ding-2018-db100-cmplx}, is a subset of the DBPedia knowledge graph. This knowledge graph is much larger in scale than FB15k-237 and WN18RR.

More details about these knowledge graphs and some relevant statistics can be found in Table \ref{tab:kgstats}.

% \subsection{Training and Evalution Setup}

% We follow the following protocol for all evaluating Neighbors' Loss on all the models. First we have obtained source codes for all the models, open-sourced by the respective authors. As the original models were kept fully intact there was no change in the evaluation methods for any of the models. 

\subsection{Hyperparameters}
To compare our results with the best performing models, we did hyperparameter tuning for each of the vanilla models. We first chose the hyperparameters reported on their respective webpage or paper and attempted to reproduce the performance of the model without any modification in the sampling algorithm. If the performance is below the published ones, only then we performed a hyperparameter tuning to search for the best model configuration. To enable fair comparison, while searching for the best hyperparameters for models with any minibatch selection algorithm, the embedding dimensions were kept constant at the value found for the best vanilla models.

% Then we fix all the hyperparameters except batch size and learning rate to train the model with Neighbors' Loss. We don't change any hyperparameter related to model architecture such as embedding dimension etc. that may change the capacity of the model. 

\subsection{Results}

%To measure the effectiveness of the proposed minibatch sampling algorithms, we compared the performance of each model trained with various proposed minbatch sampling algorithms. 
We evaluate our methods on the datasets FB15k-237, WN18RR and DB100K for all the baseline models listed in Section \ref{sec:background}. The results are summarized in Table \ref{tab:results-all}. We use the standard evaluation measures for the task -- Mean Reciprocal Rank (MRR), Mean Rank (MR), Hits@k for k = 1, 3 and 10.

The results for applying our sampling method for training the models for the three different knowledge graphs are quite different. For DB100K, we observed major improvements across all different metrics for the RotatE model, obtaining significantly higher numbers than the existing state-of-the-art (improvements of ~33\% for MRR, ~26\% for Hits@3, and so on). For FB15k-237, improvements were minor and in MRR, Hits@1 and Hits@3 only. But for WN18RR, model performance remained almost the same. 

% From the results, it can observed the FB15k-237 has consistent improvements across the models, whereas for WN18RR only with TransE and TuckER models we observe improvement. 

We suspect that the disparity in performance across datasets is caused by varying sparsity and scale of the knowledge graphs. DB100k's average and median degree is 12 and 7 (see Table \ref{tab:kgstats}) compared to 37.4 and 22 for FB15k-237. Also as pointed out by \cite{kbat}, the hierarchical nature of the WordNet graph can be more challenging for certain models. 

% In the table \ref{tab:fb15k237-results} we have reported the reproduced results for the models with their original loss and the results after making the modification.

% We would also like to point out that the baseline results for DB100K dataset with vanilla ComplEx model as reported in \cite{ding-2018-db100-cmplx} was improved by nearly a 12\% (absolute) improvement in MRR simply by doing an hyperparameter search. We report both the baselines in Table \ref{tab:results-all} for reference.

% Aside from that, the model's experssiveness is also a key factor for deciding which model is more suitable for different types of knowledge graphs .

% TODO:
% Talk about how for FB15k-237 consistent improvements were observed but
% for WN18RR only for certain models improvements were seen.

\section{Conclusions}
In Deep Learning, minibatch stochastic gradient descent is a key concept for training deep neural networks and it is used for almost every model published in the last decade. Although, the same is also true for Knowledge Graph Completion models, we analytically find the random selection based method for minibatch sampling lacking in terms of expected degree of entities, $E[D]$. For KGC models, this parameter directly translates to how good the estimate of the parameter updates are. We show that simple algorithms for increasing occurrence of entities in a minibatch do not work for graph structured data. Hence, we proposed different random walk based methods for sampling minibatches and showed that these minibatch samples have better connectivity properties than the randomly sampled minibatches.

While we obtain large improvements for DB100k dataset, the improvements across datasets and methods were not consistent. We identify two possible reasons for this. First, most models for the KG completion task are not convex in nature. So, usual results (and intuitions) about loss convergence and solutions found by the optimization method might not hold true. Second reason is that the candidate knowledge graphs used for analysis differ from each other in so many qualities that it is hard to point out one particular reason responsible for the observed behaviour of the models. All said, our proposed method achieves the state-of-the-art performance on the DB100K dataset, showing huge improvements over the RotatE model. Also because of the simplicity of the proposed sampling technique, it may be applied in any future models for KGC that might be developed by researchers.

\bibliographystyle{named}
\bibliography{main}

\end{document}